\documentclass[reprint,prb,amsmath,amssymb,amsfonts,showkeys]{revtex4-2}
\usepackage{graphicx}
\usepackage{dcolumn}
\usepackage{bm}
\usepackage{txfonts}
\usepackage{multirow}

\begin{document}

\title{Crossover from itinerant to localized states in the thermoelectric oxide [Ca$_2$CoO$_3$]$_{0.62}$[CoO$_2$]}

\author{H.~Sakabayashi}
\author{R.~Okazaki}
\affiliation{Department of Physics, Faculty of Science and Technology, Tokyo University of Science, Noda 278-8510, Japan}

\begin{abstract}

The layered cobaltite [Ca$_2$CoO$_3$]$_{0.62}$[CoO$_2$], often expressed as the approximate formula Ca$_3$Co$_4$O$_9$,
is a promising candidate for efficient oxide thermoelectrics
but an origin of its unusual thermoelectric transport is still in debate.
Here 
we investigate  \textit{in-plane} anisotropy of the transport properties in a broad temperature range to examine 
the detailed conduction mechanism.
The in-plane anisotropy between $a$ and $b$ axes is clearly observed both in the resistivity and the thermopower,
which is qualitatively understood with a simple band structure of the triangular lattice of Co ions 
derived from the angle-resolved photoemission spectroscopy experiments.
On the other hand, at high temperatures, the anisotropy becomes smaller and the resistivity 
shows a temperature-independent behavior, both of which indicate a hopping conduction of localized carriers.
Thus the present observations reveal a crossover from low-temperature itinerant to high-temperature localized states,
signifying both characters for the enhanced thermopower.

\end{abstract}

\maketitle

\section{Introduction}

The layered cobaltites provide fascinating platform for exploring functional properties of oxides
both in fundamental and applicational viewpoints \cite{Schipper2017}.
Since the discovery of large thermopower in the metallic Na$_x$CoO$_2$ \cite{Terasaki1997},
this class of materials has been recognized as potential oxide thermoelectrics
that possesses a high-temperature stability in air  \cite{Snyder2008,Koumoto2010,Hebert2016}.
The crystal structure of this family is composed of 
two subsystems of an insulating layer and 
CdI$_2$-type CoO$_2$ conduction layer 
alternately stacked
along the $c$ axis.
As for the origin of the large thermopower coexisting with the metallic conductivity,
Koshibae \textit{et al.} have suggested a localized model in which  
a hopping conduction of correlated $d$ electrons
with spin and orbital degeneracies involves large entropy flow \cite{Koshibae2000}.
This picture is supported by magnetic field dependence of the thermopower \cite{Wang2003} and 
 is also discussed in semiconducting cobalt oxides \cite{Takahashi2018}.
On the other hand, 
Kuroki \textit{et al.} have proposed an itinerant model based on 
``pudding-mold'' band structure, in which the difference in velocities of electrons and holes 
is crucial \cite{Kuroki2007JPSJ}. 
Indeed, such a peculiar band shape is 
observed by angle-resolved photoemission spectroscopy (ARPES) experiments \cite{Yang2004,Yang2005,Qian2006,Geck2007,Arakane2011} 
and 
a large value of thermopower is calculated accordingly \cite{Hamada2007,Chen2017},
remaining the detailed conduction mechanism in this system controversial.

The misfit oxide [Ca$_2$CoO$_3$]$_{0.62}$[CoO$_2$] \cite{Funahashi2000,Masset2000,Miyazaki2002}
is a suitable compound to shed light on above fundamental issue,
because it straddles a border between localized and itinerant states 
of Co $3d$ electrons,
from which interesting emergent phenomena appear in correlated electron systems \cite{Dagotto2005}.
As shown in Figs. 1(a) and 1(b), this material has a rocksalt-type Ca$_2$CoO$_3$ block as an insulating layer
and its $b$-axis lattice parameter $b_2$ is different from that of CoO$_2$ layer $b_1$,
leading to a misfit structure with incommensurate 
ratio of $b_1/b_2\simeq0.62$
while
this system is often referred as
the approximate formula 
Ca$_3$Co$_{4}$O$_9$. 
The metallic transport properties,
along with 
a spin-density-wave (SDW) formation at $T_{\rm SDW}\simeq 30$~K \cite{Sugiyama2002,Sugiyama2003,Murashige2017}, 
indicate an itinerant nature,
although the magnetic structure revealed by recent neutron experiments is quite unconventional \cite{Ahad2020_2}.
On the other hand, 
compared with that of Na$_x$CoO$_2$,
this compound has moderately high resistivity \cite{Mikami2006},
which is close to the Ioffe-Regel limit \cite{Masset2000}.
Negative magnetothermopower is also found in [Ca$_2$CoO$_3$]$_{0.62}$[CoO$_2$] \cite{Limelette2006}.
In addition, 
large 
thermopower of 
$Q\simeq 130$~$\mu$V/K
near room temperature is well explained in the extended Heikes formula
based on the localized hopping picture of correlated electrons \cite{Klie2012},
which prevails not only in the conduction layer but also the rocksalt one
as suggested by recent  spectroscopic studies \cite{Ahad2020}.
These experimental facts imply
a complicated coexistence of itinerant and localized nature in this compound.

\begin{figure}[b]
\begin{center}
\includegraphics[width=1\linewidth]{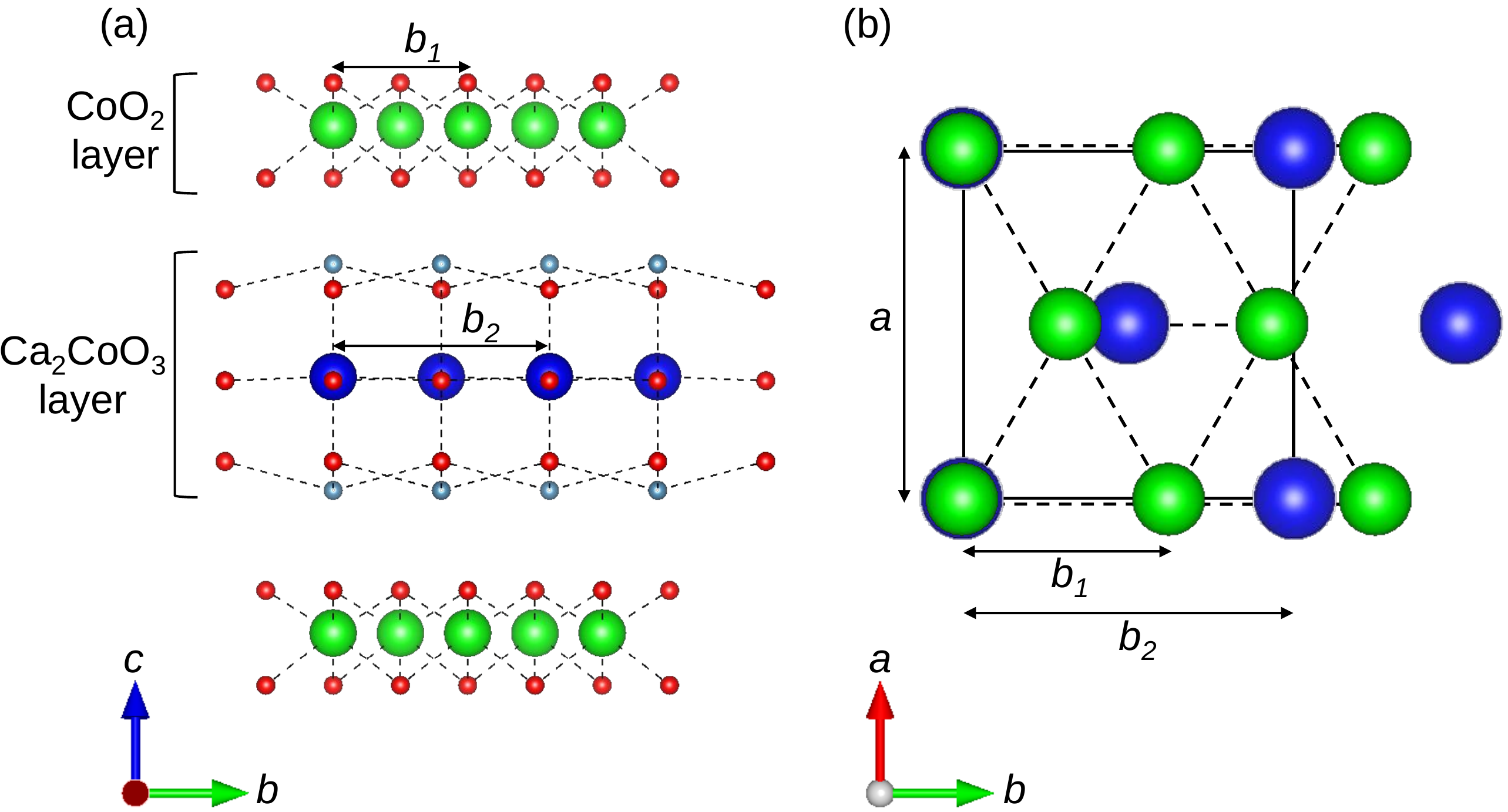}
\caption{
Schematic view of the crystal structure of layered [Ca$_2$CoO$_3$]$_{0.62}$[CoO$_2$]
projected from (a) $a$ axis and (b) $c$ axis drawn by VESTA \cite{vesta}.
While the $a$-axis lattice parameter $a$ is common,
the $b$-axis lattice parameter of the rocksalt layer $b_2$ is different from that of CoO$_2$ layer $b_1$,
resulting in a misfit structure with incommensurate 
ratio of $b_1/b_2\simeq0.62$.
The Co ions in CoO$_2$ (green) and rocksalt (blue) layers are shown in different colors for clarity.}
\end{center}
\end{figure}

In this study, we argue the itinerancy and localization of conduction electrons in [Ca$_2$CoO$_3$]$_{0.62}$[CoO$_2$]
by means of \textit{in-plane} transport anisotropy measurements between $a$- and $b$-axis directions.
This is less investigated so far compared to the strong anisotropy between in-plane and out-of-plane directions \cite{Masset2000,Tang2001}
but is essential for thorough understanding of the underlying conduction mechanism. 
We find a considerable temperature dependence of the in-plane anisotropy
both in resistivity and thermopower.
Below room temperature, the in-plane anisotropy is relatively large and qualitatively explained by the 
anisotropy of the electronic velocities near the Fermi energy estimated from the results of ARPES experiments \cite{Takeuchi2005}.
This is also consistent with the results of recent band calculation \cite{Lemal2017}, indicating the itinerant nature in this system.
Near $T_{\rm SDW}\simeq 30$~K, the in-plane anisotropy drastically varies 
possibly due to a reconstruction of the Fermi surface.
Above room temperature, in contrast, we find that 
the in-plane anisotropy is close to unity as temperature increases,
which is captured as a localized picture.
The present results unveil the temperature-induced crossover from itinerant to localized states
in [Ca$_2$CoO$_3$]$_{0.62}$[CoO$_2$].

\section{Experiments}

\begin{figure}[t]
\begin{center}
\includegraphics[width=1\linewidth]{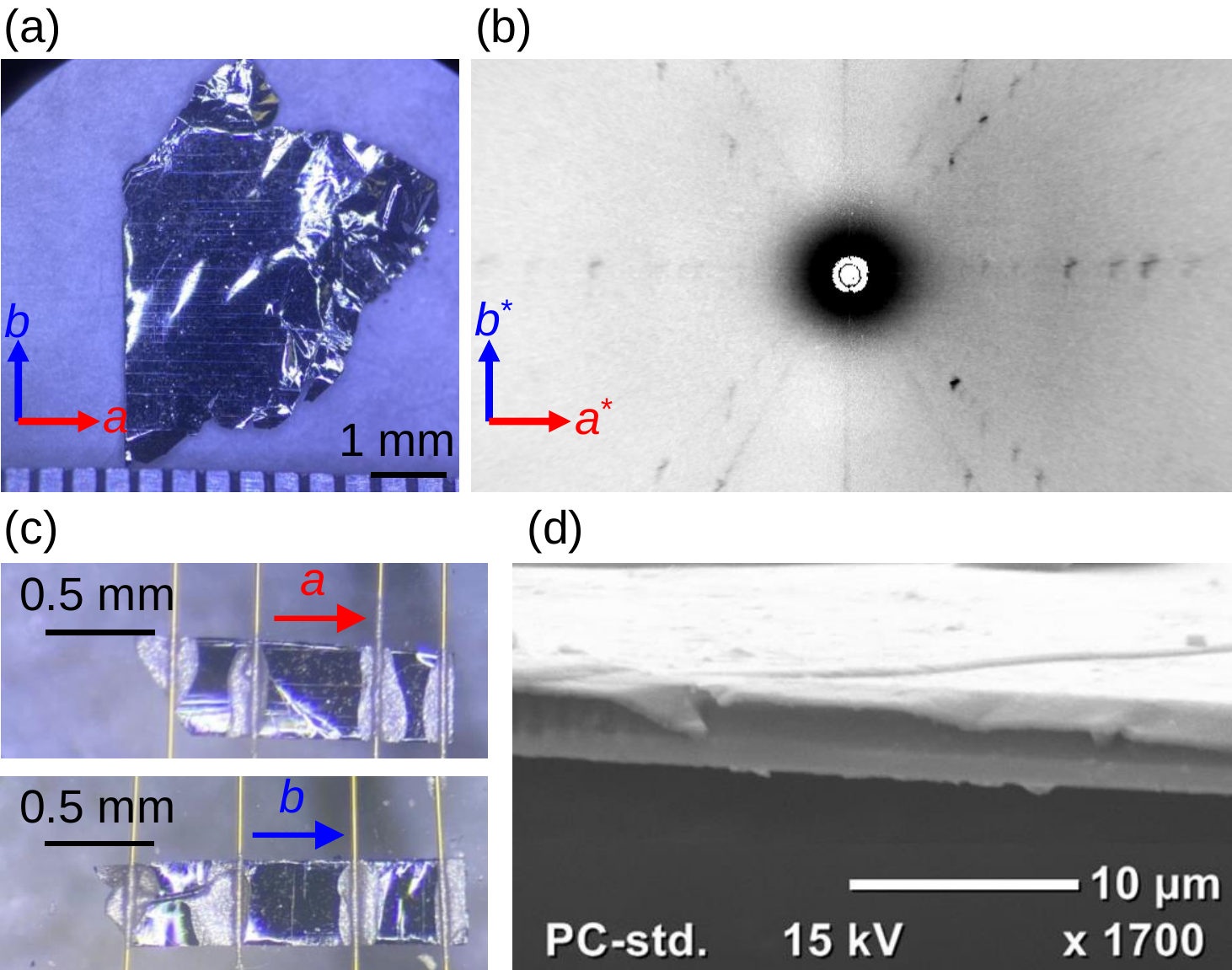}
\caption{
(a) Photograph of a single crystal of [Ca$_2$CoO$_3$]$_{0.62}$[CoO$_2$] and 
(b) Laue pattern.
(c) Photographs of samples for the resistivity measurements. Two samples with the rectangular shape of $\sim1\times 0.3$~mm$^2$
were obtained by cutting one single crystal.
(d) SEM image of [Ca$_2$CoO$_3$]$_{0.62}$[CoO$_2$] for the sample thickness measurement.
}
\end{center}
\end{figure}

Single crystals of
[Ca$_2$CoO$_3$]$_{0.62}$[CoO$_2$] were
grown by a flux method \cite{Ikeda2016}. 
Powders of CaCO$_3$ (99.9\%) and Co$_3$O$_4$ (99.9\%)
were mixed in a stoichiometric ratio and
calcined two times in air at 1173~K for 20~h with intermediate grindings.
Then KCl  (99.999\%) and K$_2$CO$_3$  (99.999\%) powders mixed with a molar ratio of $1:4$ was added 
with the calcined powder as a flux. 
The concentration of [Ca$_2$CoO$_3$]$_{0.62}$[CoO$_2$] was set to be 1.5\% in molar ratio. 
The mixture was put in an alumina crucible and 
heated up to 1123~K in air with a heating rate of 200~K/h.
After keeping 1123~K  for 1~h, 
it was slowly cooled down  with a rate of 1~K/h, 
and at 1023~K, the power of the furnace was switched off.
As-grown samples were rinsed in distilled water to remove the flux and then
annealed in air at 573~K for 3~h.

The typical dimension of obtained single crystals is $\simeq 4\times 4\times0.01$~mm$^3$ 
as shown in Fig. 2(a).
The crystal orientation was determined by the Laue method.
Although the spot intensity is weak in such thin samples,
three-fold symmetry from CoO$_2$ layer and four-fold symmetry 
from the rocksalt layer are resolved in the Laue pattern shown in Fig. 2(b).
To discuss the anisotropy precisely, we cut one single crystal into two samples with the
same rectangular shape of $\simeq 1\times0.3$~mm$^2$ 
for the transport measurements along the $a$ and $b$ axes
as shown in Fig. 2(c).
This method enables us to compare the transport properties 
of the samples with the same oxygen contents.
Note that the resistivity anisotropy was also checked by utilizing the Montgomery method near room temperature \cite{Santos2011},
although it may produce a fairly large systematic error bar \cite{Tanatar2007}.
The sample thickness of $\simeq 5$~$\mu$m was determined by the scanning electron microscopy (SEM)
as shown in Fig. 2(d).
The resistivity and the thermopower were simultaneously measured by using a conventional four-probe method and a steady-state method, respectively. 
The thermoelectric voltage from the wire leads was carefully subtracted. 
We used a Gifford-McMahon refrigerator below room temperature and an electrical furnace for high-temperature measurement.

\section{Results and discussion}

\begin{figure}[t]
\begin{center}
\includegraphics[width=0.9\linewidth]{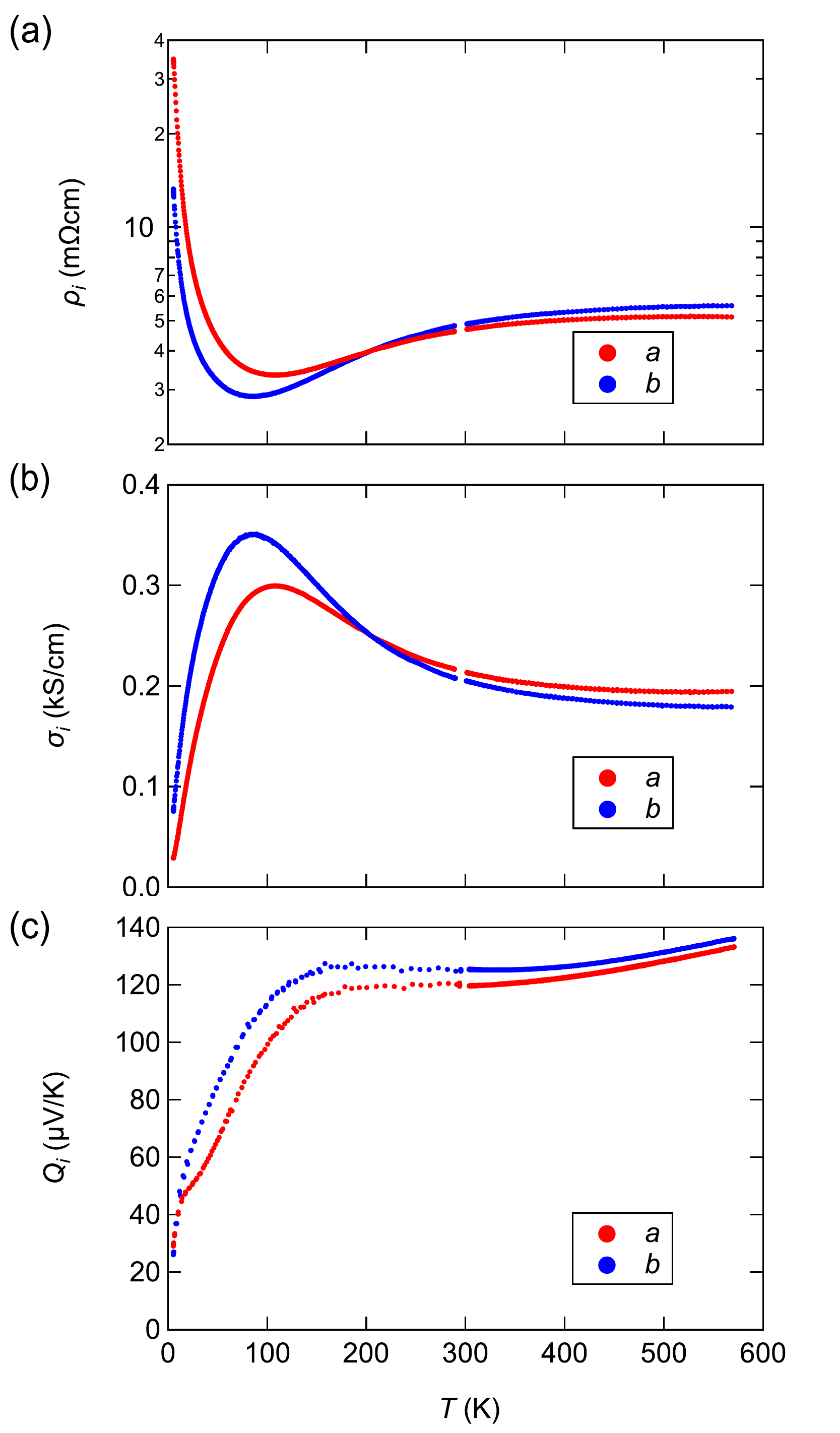}
\caption{
Temperature variations of 
(a) resistivity $\rho_i$ ($i=a,b$), (b) conductivity $\sigma_i=\rho_i^{-1}$, and 
(c) thermopower $Q_i$ measured along the $a$- (red) and $b$-axis (blue) directions
in [Ca$_2$CoO$_3$]$_{0.62}$[CoO$_2$].
}
\end{center}
\end{figure}

\subsection{Temperature variations of resistivity and thermopower}

Figures 3 summarize the temperature variations of 
the in-plane transport properties in [Ca$_2$CoO$_3$]$_{0.62}$[CoO$_2$].
Hereafter we use an abbreviated form like $\rho_a (=\rho_{aa})$ for the resistivity measured along the $a$-axis direction.
Overall behaviors are well reproduced compared with the earlier reports in which the 
transport coefficients are measured with no distinction among the in-plane directions \cite{Masset2000}.
In addition, the thermopower measured along the $b$ axis $Q_b$ is larger than 
that along the $a$ axis $Q_a$, consistent with recent theoretical calculation \cite{Lemal2017}.
At low temperature below $\sim100$~K, 
the resistivity shows an insulating behavior while the thermopower seems to be metallic, 
which are discussed in terms of carrier localization \cite{Bhaskar2014}, pseudogap opening \cite{Hsieh2014}, or quantum criticality \cite{Limelette2010}.
Near room temperature, 
the thermopower shows a temperature-independent behavior with 
a relatively large value of $Q\simeq 130$~$\mu$V/K,
quantitatively explained by the 
extended Heikes formula  of
\begin{align}
Q=-\frac{k_B}{e}\ln\left(\frac{g_3}{g_4}\frac{y}{1-y}
\right),
\label{exHeikes}
\end{align} 
where $k_B$ is the Boltzmann coefficient, $e$ the elementary charge,
$g_3$ and $g_4$ the spin and orbital degeneracies of Co$^{3+}$ and Co$^{4+}$ ions, 
respectively, and
$y$ the Co$^{4+}$ (hole) concentration \cite{Koshibae2000}.
Enhancement of the thermopower above room temperature may be possibly
attributed to a small spin-state change near $T\simeq 380$~K \cite{Wu2012},
above which the degeneracy ratio $g_3/g_4$ may vary with temperature.
At this temperature, a small anomaly has been observed in several quantities such as the resistivity, magnetic susceptibility, heat capacity, and lattice constants \cite{Masset2000,Sugiyama2002,Wu2012}.
On the other hand, the magnitude of the resistive anomaly may be sample-dependent \cite{Hejt2015} and is not resolved in the present samples.
Although the present measurements are limited below 600~K, 
the increase of thermopower may continue up to 1000~K 
according to high-temperature transport experiments in this compound \cite{Shikano2003}.

\begin{figure}[t]
\begin{center}
\includegraphics[width=0.8\linewidth]{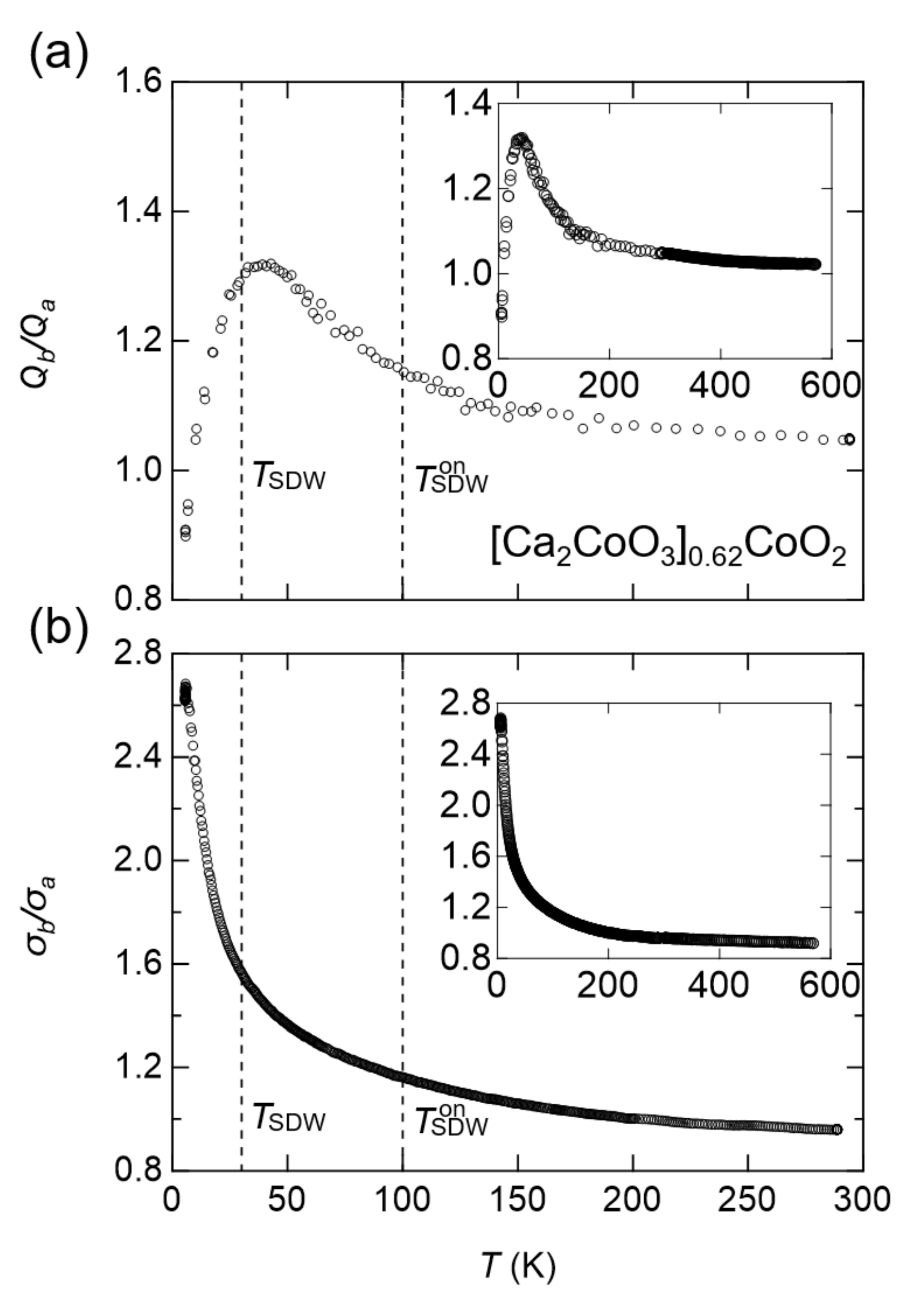}
\caption{
Temperature variations of 
the transport anisotropy in (a) thermopower $Q_b/Q_a$ and (b) electrical conductivity 
$\sigma_b/\sigma_a$ below room temperature. The SDW transition temperature $T_{\rm SDW}\simeq 30$~K and its onset 
$T_{\rm SDW}^{\rm ON}\simeq 100$~K are indicated by the dashed lines.
The insets depict these anisotropies in the whole temperature range measured in the present study.}
\end{center}
\end{figure}

\subsection{Localized state at high temperatures}

We first discuss  high-temperature transport.
The inset of Fig. 4(a) shows the temperature dependence of  
the anisotropy of  thermopower $Q_b/Q_a$.
The anisotropy $Q_b/Q_a$ decreases with 
increasing temperature and becomes close to unity near 600~K. 
This behavior is well consistent with the localized model, 
because the thermopower at high temperatures can be expressed by using 
 chemical potential $\mu$ as
\begin{align}
Q=-\frac{\mu}{eT},
\label{Q_highT}
\end{align} 
which leads to the Heikes formula of Eq. (\ref{exHeikes})  \cite{Koshibae2000},
and 
the chemical potential in the numerator
is a \textit{thermodynamic} quantity, which does not give the anisotropic property.
Note that other parameters to produce the anisotropy such as  velocity and  relaxation time are cancelled out in Eq. (\ref{Q_highT}).
Therefore, the thermopower anisotropy, 
which should be unity within the Heikes formula, 
is an indicator for localized electronic state.

Moreover, as shown in Fig. 3(a), 
the resistivity becomes less temperature-dependent in both directions at high temperatures.
Such a behavior has also been observed in high-temperature transport study while the in-plane orientation is undetermined \cite{Shikano2003},
and is described as the hopping conduction at the Ioffe-Regel limit \cite{Ioffe1960,Gurvitch1981,Hussey2004},
at which the Fermi wavelength $\lambda_F(=2\pi/k_F)$ ($k_F$ being  Fermi wavenumber) is 
comparable to the mean free path $l$ of conduction electrons.
Indeed, the dimensionless value of $k_Fl$ estimated from 
\begin{align}
k_Fl=\frac{2\pi\hbar c_0}{e^2\rho},
\label{kFl_estimation}
\end{align} 
where $\hbar$ is the reduced Planck constant and $c_0$ is the $c$-axis length \cite{Hsieh2014}, is calculated as
$k_Fl\sim 0.54$ for the $a$- and $0.50$ for the $b$-axis directions.
Although there is an unavoidable error bar mainly due to the sample thickness,
these values are close to unity, 
indicating the hopping conduction.
Thus, these transport coefficients indicate the localized nature at high temperatures.

\subsection{Itinerant state at low temperatures}

We then focus on low-temperature transport anisotropy below room temperature.
The main panels of Figs. 4(a) and 4(b) show the temperature variations of 
the anisotropies of thermopower $Q_b/Q_a$ and  electrical conductivity 
$\sigma_b/\sigma_a$ below room temperature, respectively.
In temperature range of 40~K $\lesssim  T \lesssim$ 300~K,
both $Q_b/Q_a$ and $\sigma_b/\sigma_a$ increase with lowering temperature.
Note that similar behaviors in thermopower and conductivity are also observed in
the related layered cobalitite (Bi,Pb)$_2$Sr$_2$Co$_2$O$_y$ \cite{Fujii2006},
implying a universal anisotropic property in the CoO$_2$-based materials
as is discussed below.

\begin{figure}[t]
\begin{center}
\includegraphics[width=0.8\linewidth]{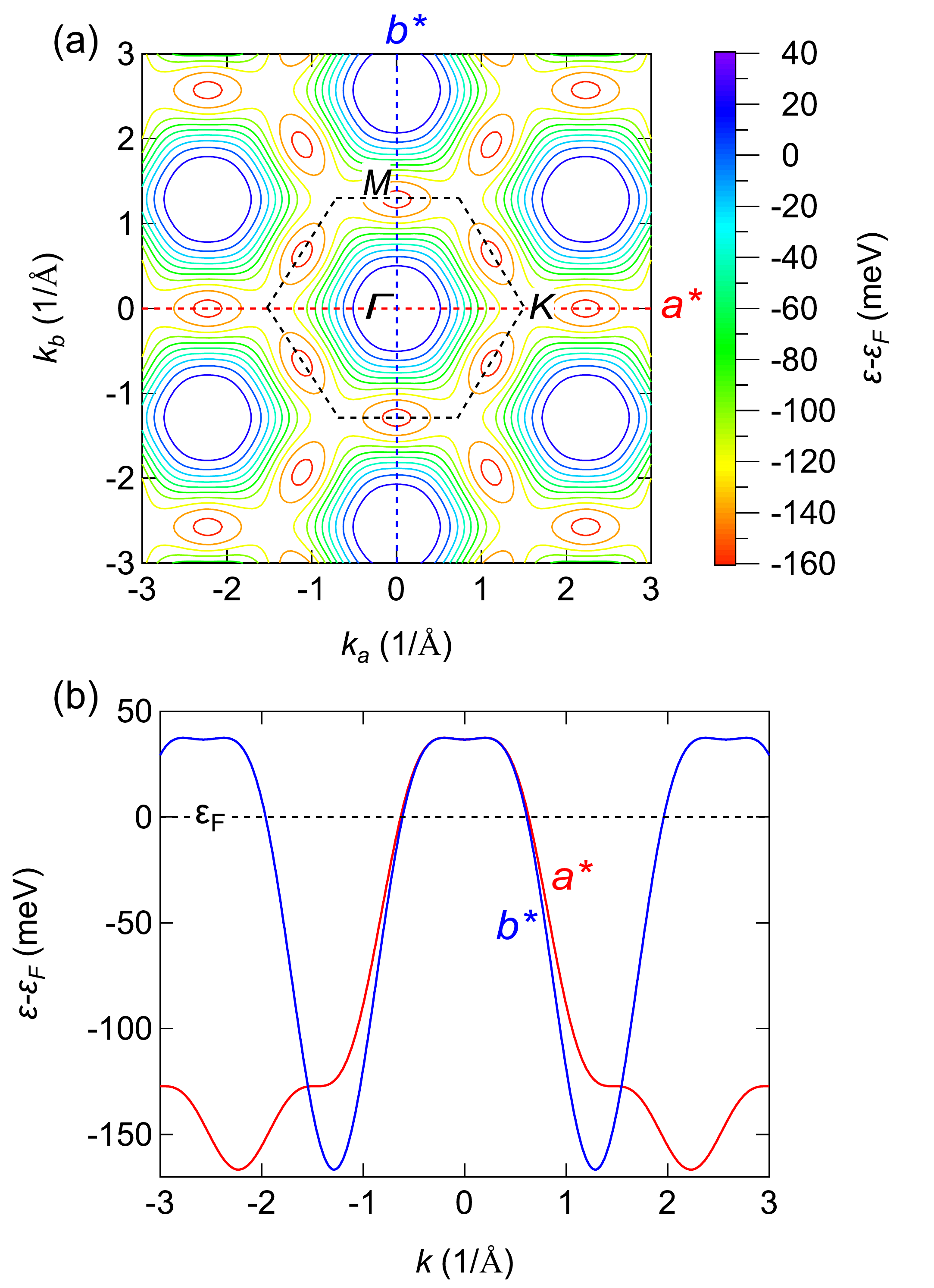}
\caption{
(a) Constant-energy surfaces near the Fermi energy calculated by using the tight-binding model [Eq. (\ref{band_hex})].
(b) Dispersion relations along $\Gamma\to{\rm K}$ ($a^*$ direction, red) and 
$\Gamma\to{\rm M}$ ($b^*$ direction, blue).
}
\end{center}
\end{figure}

To clarify the origin of the anisotropic transport at low temperatures,
we discuss the electronic band structure measured by ARPES at $T=40$~K \cite{Takeuchi2005}.
Note that the band structure is also proposed theoretically but 
there is a difficulty originating from the misfit structure of this system,
in which an approximate formula is required for calculations \cite{Lemal2017,Asahi2002,Soret2012,Rebola2012,Amin2017}.
In fact, the calculated thermopower strongly depends on the calculation methods \cite{Lemal2017,Amin2017}.
Thus, the present results may offer an experimental clue for the challenging issue to theoretically obtain the transport properties precisely in correlated electron systems with an incommensurate structure like this material.
Now the band structure experimentally obtained in [Ca$_2$CoO$_3$]$_{0.62}$[CoO$_2$]
is well fitted by 
a tight-binding dispersion relation for the hexagonal lattice with 
primitive vectors $\vec{a_h}$ and $\vec{b_h}$ as \cite{Powell}
\begin{align}
\varepsilon({\bm k}) &= \varepsilon_0-2t\left\{\cos(k_aa_h)+2\cos\left(\frac{\sqrt{3}}{2}k_ba_h\right)
\cos\left(\frac{k_aa_h}{2}\right)\right\}\nonumber \\
& -2t'\left\{\cos(2k_aa_h)+2\cos\left(\sqrt{3}k_ba_h\right)
\cos\left(k_aa_h\right)\right\},
\label{band_hex}
\end{align} 
where $\varepsilon_0 = -72.6$ meV is a constant and $a_h = 2.82$~\AA~the lattice constant.
In this model, the conducting CoO$_2$ layer is considered only,
and modeled as a hexagonal lattice.
$t = -25.4$ meV and $t' = 7.2$ meV denote the transfer integrals
between atomic orbitals connected with $\vec{a_h}$ and those with $2\vec{a_h}$.
As is discussed in Ref. \onlinecite{Takeuchi2005}, 
tight-binding fit with $t''$, which is transfer integral along $\vec{a_h}+\vec{b_h}$ direction,
leads to wrong result probably due to a one-dimensional $\sigma$-bond formation along $\vec{a_h}$ direction.
Figure 5(a) shows the calculated contour plot for the constant-energy surfaces, 
in which a Fermi surface with a hexagonal shape is 
confirmed as seen in ARPES experiments \cite{Takeuchi2005}.
Note that $k_a$ and $k_b$ directions correspond to $\sim a$ and $b$ directions in real space, respectively \cite{axis}.

\begin{figure}[b]
\begin{center}
\includegraphics[width=1\linewidth]{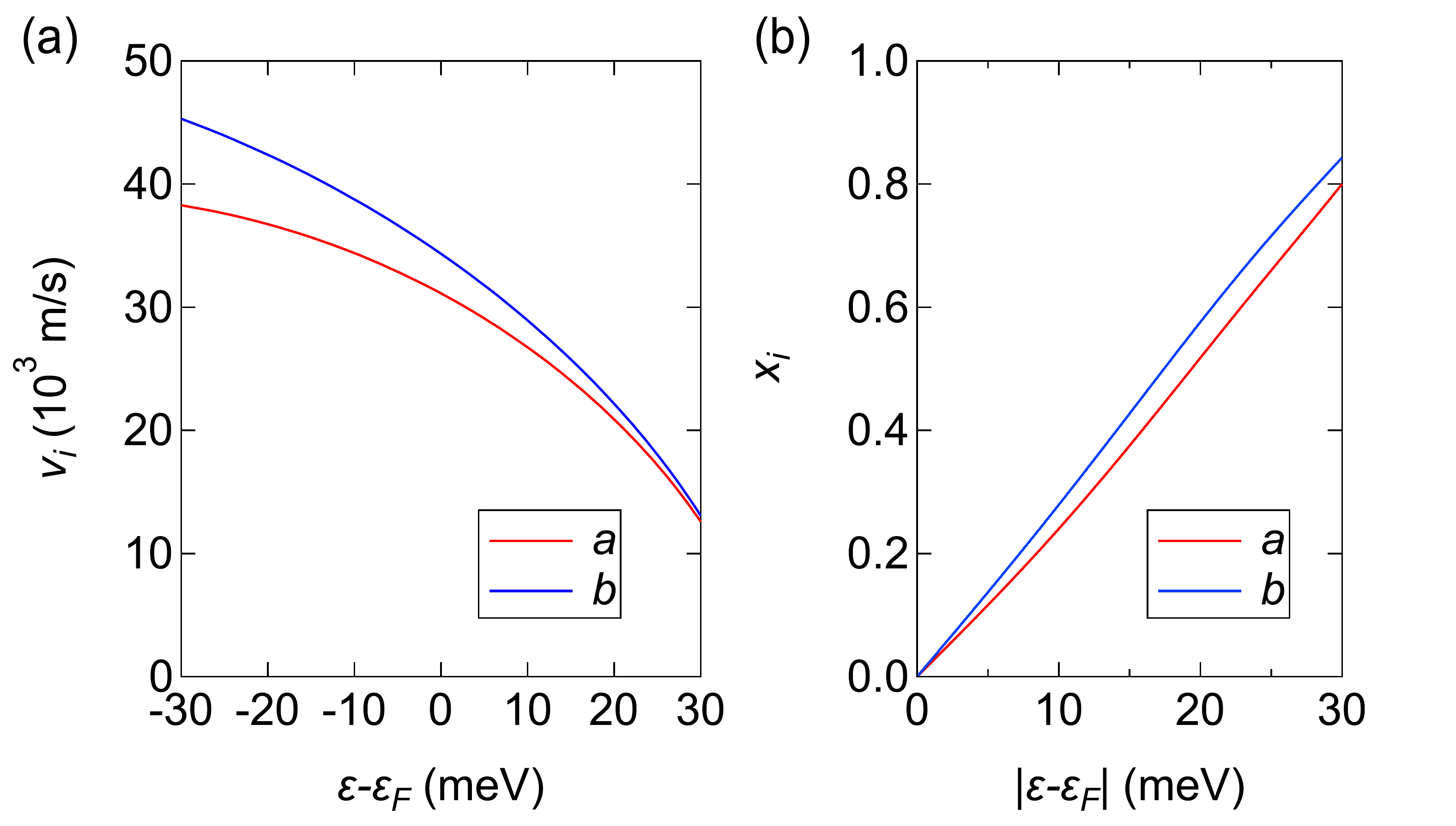}
\caption{
(a) Electronic velocity $v_i = 1/\hbar(\partial \varepsilon/\partial k_i)$ as a function of 
energy measured from the Fermi energy for $a$ (red) and $b$ (blue) directions.
The velocity $v_a$ ($v_b$) is calculated for $k_b = 0$ ($k_a=0$).
(b) A velocity ratio $x_i \equiv (v^2_{i,h}-v^2_{i,e})/(v^2_{i,h}+v^2_{i,e})$ for $a$ (red) and $b$ (blue) directions
as a function of the energy measured from $\varepsilon_F$ in magnitude.}

\end{center}
\end{figure}

Figure 5(b) shows the dispersion relations along the $\Gamma\to{\rm K}$ and 
$\Gamma\to{\rm M}$ directions.
The flat region in the top of these dispersions indicates a pudding mold energy band,
in which the thermopower is approximately determined by the 
difference in the velocities of electrons and holes as \cite{Kuroki2007JPSJ}
\begin{align}
Q_i = \frac{k_B}{e}\frac{\sum(v^2_{i,h}-v^2_{i,e})}{\sum(v^2_{i,h}+v^2_{i,e})},
\label{pudding_Q}
\end{align} 
where $v_{i,e}$ and $v_{i,h}$
are the velocities of electrons and hole  for the $i$ direction ($i=a,b$),
respectively.
Figure 6(a) presents the carrier velocity
calculated as 
$v_a = 1/\hbar(\partial \varepsilon/\partial k_a)|_{k_b=0}$ 
and
$v_b = 1/\hbar(\partial \varepsilon/\partial k_b)|_{k_a=0}$ 
as a function of 
energy near the Fermi energy.
Although 
the summation in Eq. (\ref{pudding_Q}) is taken over all the ${\bm k}$ states within the thermal energy $k_BT$,
for simplicity we only consider these $v_a$ and $v_b$, which largely contribute to the transport coefficients for each direction.
In this approximation, one obtains $v_{i,e} = v_i(\varepsilon>\varepsilon_F)$ and $v_{i,h} = v_i(\varepsilon<\varepsilon_F)$.
Figure 6(b) depicts a velocity ratio $x_i \equiv (v^2_{i,h}-v^2_{i,e})/(v^2_{i,h}+v^2_{i,e})$ for each direction
as a function of 
the energy from the Fermi energy in magnitude.
In the present energy range, 
which is comparable to the thermal energy $k_BT$, 
$x_b>x_a$ holds, leading to $Q_b>Q_a$.
This is consistent with the present result shown in Fig. 3(c),
indicating a validity of itinerant picture based on the observed Fermi surface.

We mention the anisotropy in the SDW phase below 30 K.
This material shows the SDW transition at $T_{\rm SDW}\simeq 30$~K and its 
onset temperature is $T_{\rm SDW}^{\rm ON}\simeq 100$~K.
As shown in Fig. 4(a), although no prominent feature is found at $T_{\rm SDW}^{\rm ON}$, 
we observed a cusp near $T_{\rm SDW}$ in the temperature dependence of the thermopower anisotropy.
The cusp structure in the thermopower anisotropy 
indicates that the electronic structure is drastically varied at $T_{\rm SDW}$,
as is seen in the significant change in thermopower anisotropy at the SDW transition in Fe-based superconductors \cite{Matusiak2018}.
On the other hand, the 
conductivity anisotropy $\sigma_b/\sigma_a$ is monotonically increased as temperature decreases and
shows no anomaly at $T_{\rm SDW}$ as seen in Fig. 4(b).
This result implies that, 
although the anisotropy of the velocities is  
crucial since $v_b>v_a$ leads to $\sigma_b>\sigma_a$ from
$\sigma_i \sim \sum \left(v^2_{i,h}+v^2_{i,e}\right)\tau$,
where $\tau$ is the relaxation time,
the 
conductivity anisotropy may be governed by the relaxation time,
which is cancelled out in the thermopower formula of Eq. (\ref{pudding_Q}).

\section{Summary}
To summarize, we have measured the anisotropies of the resistivity and the thermopower in
the layered [Ca$_2$CoO$_3$]$_{0.62}$[CoO$_2$] and found the considerable temperature variations.
In high-temperature range, the anisotropy in the thermopower becomes close to unity,
indicating a localized picture. On the other hand, low-temperature anisotropies are qualitatively
explained in the itinerant band picture based on the results from ARPES, and the change in the electronic structure associated with the 
SDW transition is probed as a cusp behavior of the thermopower anisotropy.
These results show a crossover from low-temperature itinerant
to high-temperature localized electronic states in this material.

\begin{acknowledgments}
The authors would like to thank S.~Makino for an early stage of the present study.
We are grateful to S.~Yoshioka for allowing us to use the scanning electron microscope (SEM).
We thank H. Yaguchi for discussion and H.~Utagawa, H.~Hatada for experimental supports.
This work was supported by JSPS KAKENHI Grants No. 17H06136, No. JP18K03503, and No. JP18K13504.
\end{acknowledgments}



\begin{thebibliography}{99}

\bibitem{Schipper2017}
F. Schipper, E. M. Erickson, C. Erk, J.-Y. Shin, F. F. Chesneau, and D. Aurbach,
J. Electrochem. Soc. {\bf 164}, A6220 (2017).

\bibitem{Terasaki1997}
I. Terasaki, Y. Sasago, and K. Uchinokura, 
Phys. Rev. B {\bf 56}, R12685 (1997).

\bibitem{Snyder2008}
G. J. Snyder and E. S. Toberer, 
Nat. Mater. {\bf 7}, 105 (2008).

\bibitem{Koumoto2010}
K. Koumoto, Y. Wang, R. Zhang, A. Kosuga, and R. Funahashi, 
Annu. Rev. Mater. Res. {\bf 40}, 363 (2010).

\bibitem{Hebert2016}
S. H\'ebert, D. Berthebaud, R. Daou, Y. Br\'eard, D. Pelloquin, E. Guilmeau, F. Gascoin, O. Lebedev, and A. Maignan,
J. Phys.: Condens. Matter {\bf 28}, 013001 (2016).

\bibitem{Koshibae2000}
W. Koshibae, K. Tsutsui, and S. Maekawa,
Phys. Rev. B {\bf 62}, 6869 (2000).

\bibitem{Wang2003}
Y. Wang, N. S. Rogado, R. J. Cava, and N. P. Ong,
Nature {\bf 423}, 22 (2003).

\bibitem{Takahashi2018}
H. Takahashi, S. Ishiwata, R. Okazaki, Y. Yasui, and I. Terasaki, 
Phys. Rev. B {\bf 98}, 024405 (2018).

\bibitem{Kuroki2007JPSJ}
K. Kuroki and R. Arita,
J. Phys. Soc. Jpn. {\bf 76}, 083707 (2007).



\bibitem{Yang2004}
H.-B. Yang, S.-C. Wang, A. K. P. Sekharan, H. Matsui, S. Souma, T. Sato, T. Takahashi, T. Takeuchi, J. C. Campuzano, R. Jin, B. C. Sales, D. Mandrus, Z. Wang, and H. Ding, 
Phys. Rev. Lett. {\bf 92}, 246403 (2004).

\bibitem{Yang2005}
H.-B. Yang, Z.-H. Pan, A. K. P. Sekharan, T. Sato, S. Souma, T. Takahashi, R. Jin, B. C. Sales, D. Mandrus, A. V. Fedorov, Z. Wang, and H. Ding, 
Phys. Rev. Lett. {\bf 95}, 146401 (2005).

\bibitem{Qian2006}
D. Qian, L. Wray, D. Hsieh, D. Wu, J. L. Luo, N. L. Wang, A. Kuprin, A. Fedorov, R. J. Cava, L. Viciu, and M. Z. Hasan, 
Phys. Rev. Lett. {\bf 96}, 046407 (2006).

\bibitem{Geck2007}
J. Geck, S. V. Borisenko, H. Berger, H. Eschrig, J. Fink, M. Knupfer, K. Koepernik, A. Koitzsch, A. A. Kordyuk, V. B. Zabolotnyy, and B. B\"uchner,
Phys. Rev. Lett. {\bf 99}, 046403 (2007).

\bibitem{Arakane2011}
T. Arakane, T. Sato, T. Takahashi, T. Fujii, and A. Asamitsu, 
New J. Phys. {\bf 13}, 043021 (2011).


\bibitem{Hamada2007}
N. Hamada, T. Imai, and H. Funashima,
J. Phys.: Condens. Matter {\bf 19}, 365221 (2007).


\bibitem{Chen2017}
S.-D. Chen, Y. He, A. Zong, Y. Zhang, M. Hashimoto, B.-B. Zhang, S.-H. Yao, Y.-B. Chen, J. Zhou, Y.-F. Chen, S.-K. Mo, Z. Hussain, D. Lu, and Z.-X. Shen,
Phys. Rev. B {\bf 96}, 081109(R) (2017).

\bibitem{Funahashi2000}
R. Funahashi, I. Matsubara, H. Ikuta, T. Takeuchi, U. Mizutani, and S. Sodeoka,
J. Appl. Phys. {\bf 39}, L1127 (2000).

\bibitem{Masset2000}
A. C. Masset, C. Michel, A. Maignan, M. Hervieu, O. Toulemonde, F. Studer, B. Raveau, and J. Hejtmanek, 
Phys. Rev. B {\bf 62}, 166 (2000).

\bibitem{Miyazaki2002}
Y. Miyazaki, M. Onoda, T. Oku, M. Kikuchi, Y. Ishii, Y. Ono, Y. Morii, and T. Kajitani,
J. Phys. Soc. Jpn. {\bf 71}, 491 (2002).

\bibitem{Dagotto2005}
E. Dagotto, Science {\bf 309}, 257 (2005).

\bibitem{vesta}
K. Momma and F. Izumi, 
J. Appl. Crystallogr. {\bf 44}, 1272 (2011).


\bibitem{Sugiyama2002}
J. Sugiyama, H. Itahara, T. Tani, J. H. Brewer, and E. J. Ansaldo, 
Phys. Rev. B {\bf 66}, 134413 (2002).

\bibitem{Sugiyama2003}
J. Sugiyama, J. H. Brewer, E. J. Ansaldo, H. Itahara, K. Dohmae, Y. Seno, C. Xia, and T. Tani,
Phys. Rev. B {\bf 68}, 134423 (2003).

\bibitem{Murashige2017}
N. Murashige, F. Takei, K. Saito, and R. Okazaki,
Phys. Rev. B {\bf 96}, 035126 (2017).

\bibitem{Ahad2020_2} 
A. Ahad, K. Gautam, K. Dey, S. S. Majid, F. Rahman, S. K. Sharma, J. A. H. Coaquira, Ivan da Silva, E. Welter, and D. K. Shukla,
Phys. Rev. B {\bf 102}, 094428 (2020).


\bibitem{Mikami2006}
M. Mikami, K. Chong, Y. Miyazaki, T. Kajitani, T. Inoue, S. Sodeoka and R. Funahashi,
J. Appl. Phys. {\bf 45}, 4131 (2006).

\bibitem{Limelette2006}
P. Limelette, S. H\'ebert, V. Hardy, R. Fr\'esard, Ch. Simon, and A. Maignan, 
Phys. Rev. Lett. {\bf 97}, 046601 (2006).


\bibitem{Klie2012}
R. F. Klie, Q. Qiao, T. Paulauskas, A. Gulec, A. Rebola, S. \"O\ifmmode \breve{g}\else \u{g}\fi{}\"ut, M. P. Prange, J. C. Idrobo, S. T. Pantelides, S. Kolesnik, B. Dabrowski, M. Ozdemir, C. Boyraz, D. Mazumdar, and A. Gupta,
Phys. Rev. Lett. {\bf 108}, 196601 (2012).

\bibitem{Ahad2020} 
A. Ahad, K. Gautam, S. S. Majid, S. Francoual, F. Rahman, F. M. F. De Groot, and D. K. Shukla,
Phys. Rev. B {\bf 101}, 220202(R) (2020).

\bibitem{Tang2001}
G. D. Tang, H. H. Guo, T. Yang, D. W. Zhang, X. N. Xu, L. Y. Wang, Z. H. Wang, H. H. Wen, Z. D. Zhang, and Y. W. Du,
Appl. Phys. Lett. {\bf 98}, 202109 (2011).

\bibitem{Takeuchi2005} 
T. Takeuchi, T. Kondo, T. Kitao, K. Soda, M. Shikano, R. Funahashi, M. Mikami, and U. Mizutani,
J. Electron Spectrosc. Relat. Phenom. {\bf 144-147}, 849 (2005).

\bibitem{Lemal2017} 
S. Lemal, J. Varignon, D. I. Bilc, and P. Ghosez,
Phys. Rev. B {\bf 95}, 075205 (2017).

\bibitem{Ikeda2016} 
Y. Ikeda, K. Saito, and R. Okazaki,
J. Appl. Phys. {\bf 119}, 225105 (2016).

\bibitem{Santos2011}
C. A. M. dos Santos, A. de Campos, M. S. da Luz, B. D. White, J. J. Neumeier, B. S. de Lima, and C. Y. Shigue,
J. Appl. Phys. {\bf 110}, 083703 (2011).

\bibitem{Tanatar2007}
M. A. Tanatar, N. Ni, G. D. Samolyuk, S. L. Budko, P. C. Canfield, and R. Prozorov,
Phys. Rev. B {\bf 79}, 134528 (2009).

\bibitem{Bhaskar2014}
A. Bhaskar, Z.-R. Lin, and C.-J. Liu, 
J. Mater. Sci. {\bf 49}, 1359 (2014).

\bibitem{Hsieh2014}
Y.-C. Hsieh, R. Okazaki, H. Taniguchi, and I. Terasaki,
J. Phys. Soc. Jpn. {\bf 83}, 054710 (2014).

\bibitem{Limelette2010}
P. Limelette, W. Saulquin, H. Muguerra, and D. Grebille,
Phys. Rev. B {\bf 81}, 115113 (2010).

\bibitem{Wu2012}
T. Wu, T. A. Tyson, H. Chen, J. Bai, H. Wang and C. Jaye,
J. Phys.: Condens. Matter {\bf 24}, 455602 (2012).

\bibitem{Hejt2015}
J. Hejtm\'anek, Z. Jir\'ak, and J.  \ifmmode \check{S}\else \v{S}\fi{}ebek,
Phys. Rev. B {\bf 92}, 125106 (2015).

\bibitem{Shikano2003}
M. Shikano and R. Funahashi, 
Appl. Phys. Lett. {\bf 82}, 1851 (2003).

\bibitem{Ioffe1960}
A. F. Ioffe and A. R. Regel,
Prog. Semicond. {\bf 4}, 237 (1960).

\bibitem{Gurvitch1981}
M. Gurvitch,
Phys. Rev. B {\bf 24}, 7404 (1981).

\bibitem{Hussey2004}
N. E. Hussey, K. Takenaka, and H. Takagi,
Philos Mag {\bf 84}, 27 (2004).

\bibitem{Fujii2006}
T. Fujii, I. Terasaki, T. Watanabe and A. Matsuda,
J. Appl. Phys. {\bf 45}, 4131 (2006).

\bibitem{Asahi2002} 
R. Asahi, J. Sugiyama, and T. Tani,
Phys. Rev. B {\bf 66}, 155103 (2002).

\bibitem{Soret2012} 
J. Soret and M.-B. Lepetit,
Phys. Rev. B {\bf 85}, 165145 (2012).

\bibitem{Rebola2012} 
A. R\'ebola, R. Klie, P. Zapol, and S. \"O\ifmmode \breve{g}\else \u{g}\fi{}\"ut,
Phys. Rev. B {\bf 85}, 155132 (2012).

\bibitem{Amin2017} 
B. Amin, U. Eckern, and U. Schwingenschl\"{o}gl,
Appl. Phys. Lett. {\bf 110}, 233505 (2017).

\bibitem{Powell} 
B. J. Powell, arXiv:0906.1640.

\bibitem{axis} 
Although the crystal system of [Ca$_2$CoO$_3$]$_{0.62}$[CoO$_2$] is monoclinic, 
the angle $\beta\sim98^{\circ}$ is close to $90^{\circ}$. We then assume that 
$a^*$ direction is close to $a$ direction. 

\bibitem{Matusiak2018} 
M. Matusiak, M. Babij, and T. Wolf,
Phys. Rev. B {\bf 97}, 100506(R) (2018).

\end{thebibliography}
\end{document}